\documentclass[aps,pre,preprint]{revtex4-2}
\usepackage{amsmath}
\usepackage{graphicx}
\usepackage{bbm}
\usepackage{subfigure}
\usepackage[caption=false]{subfig}
\begin{document}
\title{Deterministic Quantum Field Trajectories and Macroscopic Effects}

\author{M. Akbari-Moghanjoughi}
\affiliation{Faculty of Sciences, Department of Physics, Azarbaijan Shahid Madani University, 51745-406 Tabriz, Iran\\ massoud2002@yahoo.com}

\begin{abstract}
In this work the root to macroscopic quantum effects is revealed based on the quasiparticle model of collective excitations in an arbitrary degenerate electron gas. The $N$-electron quantum system is considered as $N$ streams coupled, through the Poisson's relation, which are localized in momentum space rather than electron localization in real space, assumed in ordinary many body theories. Using a new wavefunction representation, the $N+1$-coupled system is reduced to simple pseudoforce equations via quasiparticle (collective quantum) model leading to a generalized matter wave dispersion relation. It is shown that the resulting dual lengthscale de Broglie's matter wave theory predicts macroscopic quantum effects and deterministic field trajectories for charges moving in the electron gas due to the coupling of the electrostatic field to the local electron number density. It is remarked that any quantum many body system composed of large number of interacting particles acts as a dual arm device controlling the microscopic single particle effects with one hand and the macroscopic phenomena with the other. Current analysis can be further extended to include the magnetic potential and spin exchange effects. Present model can also be used to confirm macroscopic entanglement of charged particles embedded in a quantum electron fluid. 
\end{abstract}

\date{\today}

\maketitle
\newpage

\section{Introduction}

It is more than a decade that walking droplets on the surface of a periodically vibrating fluid have fascinated scientists with their many pseudoquantum features \cite{cou1,cou2}. Floating of a droplet on a vibrating bath of fluid for the first time was described by Jearl Walker in 1978 in a Scientific American amateur scientist article. The recent pioneering experiments have \cite{bush1,cou3} revealed that the macroscopic millimeter-sized walking droplets on a suitably driven fluid surface can mimic variety of quantum-like effects which were already assumed to emerge at atomic and molecular scales \cite{arn}. The most prominent interference effect has been experimentally confirmed for droplets passing through single and double slit barriers \cite{pro,cou4} which are quite reminiscent to their quantum counterparts. However, at the complete contrast to the Copenhagen interpretation, the mobile droplets are observed to pass through only one slit at a time. The later effect seems to give a credit to the initially ignored deterministic pilot wave quantum theory of de Broglie \cite{de1,de2} which was then re-established by Bohm in 1952 \cite{bohm1,bohm2}. Subsequent experiments on walking droplets have shown yet more interesting hydrodynamic quantum analog aspects such as quantum tunneling \cite{eddi}, bound state orbital quantization \cite{fort1,oza,harris} and Landau levels \cite{fort2}. Therefore, classical experiments with tiny magic droplets seem to fundamentally alter our conceptions of wave-particle duality and consequently foundations of quantum mechanics itself in near future \cite{zee}.

On the other hand, while the novel classical analogy between quantum particles and bouncing droplets may have shed light on the understanding of quantum weirdness, the underlying physical and mathematical connections between de Broglie-Bohm pilot-wave and the hydrodynamic theories is still a matter of debate \cite{chris}. It is however apparent that the akin similarities between quantum wave-particle phenomenon and the interacting droplet with its own hydrodynamic wave is pointing at a much deeper physical description of the quantum mechanics. The later suggestion has motivated an intensified research \cite{gil,hu,harris2,oza1,oza2,oza3,bush2,mil,bush3,lab,puc,burn,tam} towards detailed theoretical and experimental investigation of the dynamics of magic bouncing droplet and its interaction with the surface waves produced by the vibrating platform. Unfortunately, not much of a progress has been made towards a unified theoretical description of the phenomenon due to the very complex nature of collective interactions in fluids and the rapid temporal evolution of the droplet specific parameters. Recent studies based on the Schr\"{o}dinger-Poisson system has shown interesting features of quantum electron fluids due to electrostatic coupling of single electron excitations and electron correlations \cite{akbquant,akbheat,akbint,akbnat1}. The collective quantum excitations of electron gas has been shown to behave as if an electron possesses two distinct de broglie's wavelengths, simultaneously \cite{akbdual}. Due to such effect hot electrons in semiconductor manage to tunnel out of the plasmonic device in resonant processes \cite{akbedge,akbnat2,akbhot}. 

In this research we show that quantum mechanics is a double arm device acting at different scales at the same time. We employ a newly developed \cite{akbnew} quasiparticle approach to study collective excitations in the electron gas without the need for Madelung hydrodynamic formalism in which each electron is assumed to be a quantum stream localized in the momentum space thereby being couples to other streams through the Poisson's relation in a multistream model. We provide the quasiparticle formalism in Sec. II. In Sec. III the generalized matter wave dispersion relation is obtained. The examples of deterministic quantum field trajectories is given in Sec. IV based on dual-tone quantum theory. The macroscopic quantum effects are introduced in the electron gas in Sec. V and conclusions are drawn in Sec. VI.  

\section{Quasiparticle Representation of Collective Modes}

Collective excitations in electron gas can be effectively modeled using the quasiparticle concept. Starting from $N$ single-electron time-dependent Schr\"{o}dinger equations coupled through Poisson's relation, we have a generalized multistream electron system as follows \cite{akbnew}
\begin{subequations}\label{ms}
\begin{align}
&i\hbar \frac{{\partial {{\cal{N}}_j}({\bf{r}},t)}}{{\partial t}} = {\cal{H}}{{\cal{N}}_j}({\bf{r}},t),\\
&\Delta \phi ({\bf{r}}) = 4\pi e\left[ {\sum\limits_{j = 1}^{N} {{{{\cal{N}}_j}({\bf{r}},t){\cal{N}}_j^*({\bf{r}},t)} - {n_0}} } \right].
\end{align}
\end{subequations}
in which the subscript $j$ denotes the $j$-th stream, $n_0$ denotes the neutralizing charge number density and ${\cal{N}}_j$ represents the corresponding single-electron wavefunction. The Hamiltonian operator acting on each electron ${\cal{H}}={\cal{K}}+{\cal{U}}$ is the addition of kinetic energy operator and the potential energy in the form of ${\cal{U}}=-e\phi+\mu$, in which $\phi$ is the self consistent electrostatic potential and $\mu$ the chemical potential of the electron gas. In the quasiparticle representation the collective modes in a quantum electron gas can be studied using the the following effective Schr\"{o}dinger-Poisson system
\begin{subequations}\label{sp}
\begin{align}
&i\hbar \frac{{\partial {\cal {N}}({\bf{r}},t)}}{{\partial t}} =  - \frac{{{\hbar ^2}}}{{2m}}\Delta {\cal {N}}({\bf{r}},t) - e\phi ({\bf{r}}){\cal {N}}({\bf{r}},t) + \mu {\cal {N}}({\bf{r}},t),\\
&\Delta \phi ({\bf{r}}) = 4\pi e\left[ {|{\cal {N}}({\bf{r}},t){|^2} - {n_0}} \right].
\end{align}
\end{subequations}
where $n = {{{{\cal{N}}}({\bf{r}},t){\cal{N}}^*({\bf{r}},t)}}$ is the local electron number density defined through the quasiparticle statefunction ${\cal{N}}({\bf{r}},t) = \sum\limits_{j = 1}^N {{{\cal{N}}_j}({\bf{r}},t)}$. Note that in the quasiparticle representation electrons do not follow independent particle or mean-field assumptions but instead are coupled via the electrostatic and chemical potential in the electron gas to the Poisson's relation and appropriate equation of state (EoS), respectively. In the classical limit the chemical potential vanishes, whereas, in the fully degenerate electron gas the chemical potential becomes identical with the constant Fermi energy of the system. In the following analysis we assume that the chemical potential of the system does not vary significantly over the characteristic quantum lengthscale, for simplicity. The quasiparticle Hamiltonian can be further generalized to include the magnetic potential, electron exchange and other effects.  

\section{Generalized de Broglie's Wavelengths and Matter Wave Dispersion}

The quasiparticle model of electron gas in Eq. (\ref{sp}) has been shown to lead to dual-lengthscale theory giving rise to two distinct de Broglie wavelengths with the ordinary wavelength characterising the single-electron excitations and the other representing the collective excitations. The linearized system ($\psi^0=1,\phi^0=0,\mu^0=\mu_0$) after the variable separation leads to the following coupled pseudoforce system
\begin{subequations}\label{pf}
\begin{align}
&i\hbar\frac{{d\varphi(t)}}{{dt}} = \epsilon\varphi(t),\\
&\Delta \Psi ({\bf{r}}) + \Phi ({\bf{r}}) + 2E\Psi ({\bf{r}}) = 0,\\
&\Delta \Phi ({\bf{r}}) - \Psi ({\bf{r}}) = 0,
\end{align}
\end{subequations}
in which $\Psi({\bf{r}})=\psi({\bf{r}})/\sqrt{n_0}$ with ${{\cal{N}}}({\bf{r}},t)=\psi({\bf{r}})\varphi(t)$ is the normalized quasiparticle wavefunction and $\Phi({\bf{r}})=e\phi({\bf{r}})/E_p$ is the normalized electrostatic potential energy with $E_p=\sqrt{4\pi e^2 n_0/m}$ is the plasmon energy and the normalized energy parameter $E=(\epsilon-\mu_0)/E_p$ is the total kinetic energy of electron gas with $\epsilon = \sum\limits_{j = 1}^N {{\epsilon_j}}$ being the energy eigenvalue of quasiparticle excitations and $\epsilon_j$ the energy of $j$-th electron in the system. In this normalization scheme the space and time are normalized with respect to the plasmon length $l_p=1/k_p$ with $k_p=\sqrt{2m E_p}/\hbar$ being the plasmon wavenumber and the inverse plasmon frequency $\omega_p=E_p/\hbar$. The first equation in (\ref{pf}) leads to the trivial time-dependence solution, $\varphi(t)=\exp(i\epsilon t/\hbar)$. Linearizing the system leads to the generalized energy dispersion relation $E=k^2/2+1/2k^2$ in which the wavenumber is normalized to the plasmon wavenumber. The dispersion relation can be regarded as composed of a term charcterizing the free electron parabolic dispersion ($k^2/2$) and an extra term ($1/2k^2$) due to long range electrostatic correlations. Therefore, in the quasiparticle model, the collective excitations are characterized by two de Broglie wavelength which are actually reciprocal of each other \cite{akbdual}. Therefore, the quasiparticle model of collective electron excitation is a dual-lengthscale quantum theory. 

\begin{figure}[ptb]\label{Figure1}
\includegraphics[scale=0.75]{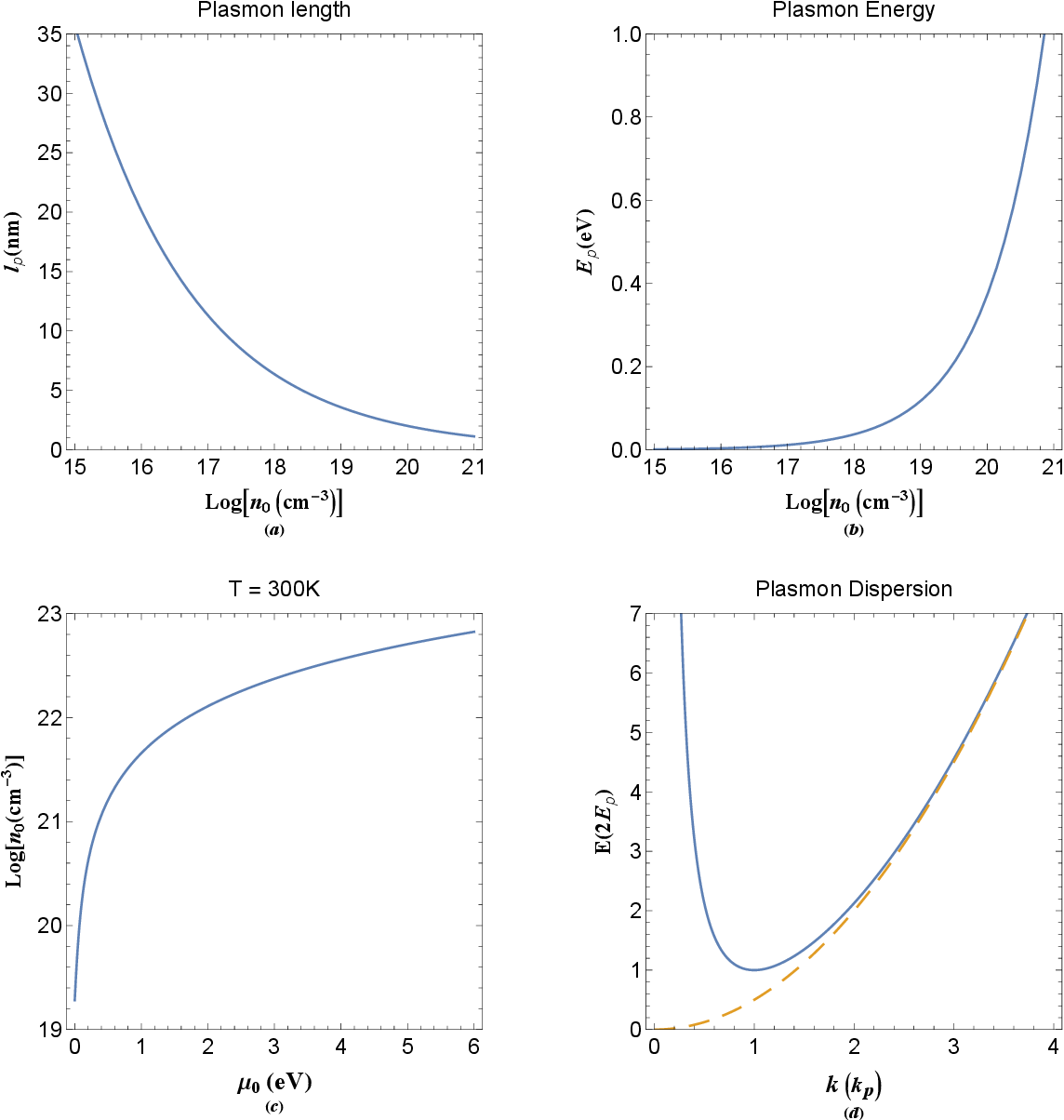}\caption{(a) The variation of characteristic plasmon length in electron gas in terms of the equilibriumm electron number density. (b) The variation of characteristic plasmon energy in electron gas in terms of the equilibriumm electron number density. (c) Variation of the chemical potential of isothermal arbitrary degenerate electron gas with respect to the equilibrium electron number density. The generalized matter wave (energy) dispersion in which energy anf wavenumber are normalized to corresponding plasmon parameters. The dashed curve in Fig. 1(d) denotes the free electron matter wave dispersion.}
\end{figure}

Figure 1 depicts the variations in different plasmon parameters and the generalized energy dispersion relation of collective excitations in interacting electron gas. Figure 1(a) shows the variations in the plasmon length in nanometer unit. It is seen that this length decreases sharply with increase in the electron gas number density dropping to few tenth of a nanometer in elemental metals. In Fig. 1(b) the plasmon energy is shown in electron Volt unit. It is remarked that this quantity increases rapidly with increase on the electron number density reaching few $eV$ in typical metals. The chemical potential of the electron gas in free electron model is related to the electron number density and quantum statistical pressure via the well-known isothermal equation of state (EoS)
\begin{subequations}\label{np1}
\begin{align}
&{n_e(\mu,T)} = \frac{{{2^{1/2}}m{^{3/2}}}}{{{\pi ^2}{\hbar ^3}}}  \int_{0}^{ + \infty } {\frac{{\sqrt{{\varepsilon}} d{\varepsilon}}}{{{e^{\beta ({\varepsilon}-\mu)}} + 1}}},\\
&{P_e(\mu,T)} = \frac{{{2^{3/2}} m{^{3/2}}}}{{3{\pi ^2}{\hbar ^3}}}\int_0^{ + \infty } {\frac{{{{\varepsilon}^{3/2}} d{\varepsilon}}}{{{e^{\beta ({\varepsilon} - {\mu})}} + 1}}.}
\end{align}
\end{subequations}
where $\beta=1/k_B T$ with $T$ being the equilibrium electron temperature and $P_e$ being the quantum statistical electron gas pressure. Variation of the chemical potential with electron number density at room temperature is depicted in Fig. 1(c). This parameter is vanishingly small for classical electron gas and is saturated in complete degeneracy limit ($n_0\simeq 10^{22}$cm$^{-3}$). The dispersion relation is shown in Fig. 1(d) in which the single-electron and collective branches connect at the quantum beating point $k=1$ \cite{akbdual}. For a given value of the quasiparticle energy (orbital) $E$ exceeding the band gap value, $E=2E_p$, the energy level intersects the dispersion curve in two de Broglie wavenumbers with the smaller one characterizing the long range electrostatic interactions and the larger one due to single electron excitations. The single-electron excitation dispersion branch is asymptotically approaches the free electron dispersion curve for higher quasiparticle energy values. It is remarked that, as the quasiparticle energy increases, the single electron and collective mode scalelengths differ significantly.   

\section{Deterministic Quantum Field Trajectory}

The pseudoforce system (\ref{pf}) in one dimension has the following simple solution \cite{akbdual}
\begin{equation}\label{wf}
\left[ {\begin{array}{*{20}{c}}
{{\Phi}(x)}\\
{{\Psi}(x)}
\end{array}} \right] = \frac{1}{{2\alpha }}\left[ {\begin{array}{*{20}{c}}
{{\Psi _0} + k_2^2{\Phi _0}}&{ - \left( {{\Psi _0} + k_1^2{\Phi _0}} \right)}\\
{ - \left( {{\Phi _0} + k_1^2{\Psi _0}} \right)}&{{\Phi _0} + k_2^2{\Psi _0}}
\end{array}} \right]\left[ {\begin{array}{*{20}{c}}
{\cos ({k_1}x)}\\
{\cos ({k_2}x)}
\end{array}} \right],
\end{equation}
in which $\Phi_0$ and $\Psi_0$ are the constants assuming $\Phi'(0)=\Psi'(0)$=0, for simplicity. The characteristic wavenumbers are given as
\begin{equation}\label{eks}
{k_1} = \sqrt {E - \alpha },\hspace{3mm}{k_2} = \sqrt {E + \alpha },\hspace{3mm}\alpha  = \sqrt {{E^2} - 1},
\end{equation}
satisfying the complementarity-like relation $k_1 k_2=1$. Note that the coupling of the electrostatic field to the wavefunction $\Psi(x)$ (the local electron density, $n$) leads to deterministic quantum trajectories for charge particles moving in the electron gas excited to given quasiparticle orbital. The normalized classical guiding equation is
\begin{equation}\label{ge}
\Gamma \frac{{{d^2}x}}{{d{t^2}}} + Q\frac{{d\Phi }}{{dx}} = 0,
\end{equation}
where $\Gamma$ and $Q$ are the particle mass and charge ratios with respect to that of electron. This is obviously different from the guiding equation in de Broglie-Bohm pilot wave theory defined through the probability density as  
\begin{equation}\label{db}
\frac{{dx}}{{dt}} = {\mathop{\rm \Im}\nolimits} {\mathop{\rm}\nolimits} \left[ {\nabla \ln \left( \Psi  \right)} \right] = \frac{{J(x)}}{{n(x)}},
\end{equation}
where $\Im$ denotes the imaginary part and 
\begin{equation}\label{j}
J(x) = \frac{i}{2}\left( {\Psi \frac{{\partial {\Psi ^*}}}{{\partial x}} - {\Psi ^*}\frac{{\partial \Psi }}{{\partial x}}} \right).
\end{equation}
The former determinism is defined through the particle-field coupling in quasiparticle model, however, the later is defined based on current density representation in Madelung fluid formalism. 

\begin{figure}[ptb]\label{Figure2}
\includegraphics[scale=0.75]{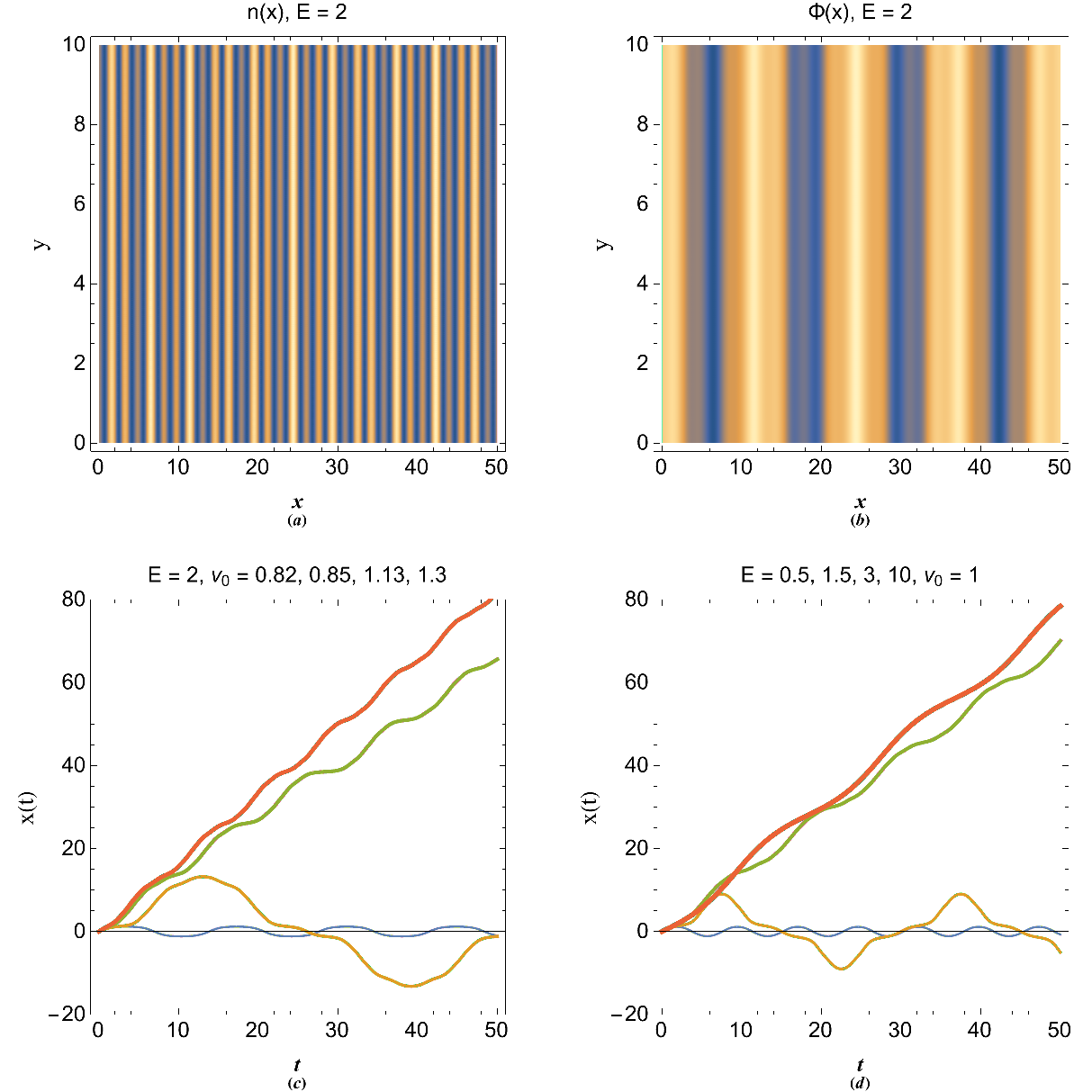}\caption{(a) Variation of local electron number density in quasiparticle orbital ($E=4E_p$). (b) Variation of electrostatic potential energy in quasiparticle orbital ($E=4E_p$). (c) Deterministic one-dimensional field trajectories of a test particle at orbital $E=2$ and different initial speeds. (d) Deterministic one-dimensional field trajectories of a test particle at different orbital and initial speed $v_0=v_p$. $v_p=(\hbar/m)k_p$ is the corresponding plasmon speed of the particle.}
\end{figure}

Figure 2(a) and 2(b) show the local electron number density and electrostatic potential energy profiles at given quasiparticle orbital ($E=2$). The electron density $n=\Psi^*\Psi$ and electrostatic potential energy variations are dual tone in nature. The later is due to two distinct de Broglie's wavenumbers governing the quasiparticle excitations in the electron gas. Figure 2(c) shows the classical charged particle trajectory in quantum-coupled electrostatic field assuming $\Gamma=Q=1$, for simplicity. It is remarked that for initial particle speed $v_0=0.82,0.85$ in a given quasiparticle orbital $E=2$ the oscillations are stationary with the particle being localized in the electron gas. However, for higher initial particle speeds the particle moves within the gas in an oscillatory path. Figure 2(d) depicts the particle trajectory for $v_0=1$ (the speed is normalized to plasmon speed $v_p=(\hbar/m) k_p$) at different quasiparticle orbital. The particle is apparently localized at low energy orbital, whereas, moving at higher quasiparticle orbital. 

\section{Macroscopic Quantum Effects}

The system of equations (\ref{pf}) has a polar solution of the form \cite{akbint}
\begin{equation}\label{mps}
\left[ {\begin{array}{*{20}{c}}
{\Phi (r)}\\
{\Psi (r)}
\end{array}} \right] = \frac{Q}{{2\alpha}r}\left[ {\begin{array}{*{20}{c}}
{{\Psi _0} + k_2^2{\Phi _0}}&{ - \left( {{\Psi _0} + k_1^2{\Phi _0}} \right)}\\
{ - \left( {{\Phi _0} + k_1^2{\Psi _0}} \right)}&{{\Phi _0} + k_2^2{\Psi _0}}
\end{array}} \right]\left( {\begin{array}{*{20}{c}}
{{{\rm{e}}^{\pm i{k_1} r}}}\\
{{{\rm{e}}^{\pm i{k_2} r}}}
\end{array}} \right),
\end{equation}
where $Q$ is the corresponding pole charge. On the other hand, in the cartesian form we have
\begin{subequations}\label{mmp}
\begin{align}
&\frac{{{\partial^2}\Psi (x,y,z)}}{{\partial{x^2}}} + \frac{{{\partial^2}\Psi (x,y,z)}}{{\partial{y^2}}} + \frac{{{\partial^2}\Psi (x,y,z)}}{{\partial{z^2}}} + \Phi (x,y,z) =  - 2E\Psi (x,y,z),\\
&\frac{{{\partial^2}\Phi (x,y,z)}}{{\partial{x^2}}} + \frac{{{\partial^2}\Phi (x,y,z)}}{{\partial{y^2}}} + \frac{{{\partial^2}\Phi (x,y,z)}}{{\partial{z^2}}} - \Psi (x,y,z) = 0,
\end{align}
\end{subequations}
where $r=\sqrt{x^2+y^2+z^2}$ and the dipole solution has the following form \cite{akbint}
\begin{subequations}\label{dp}
\begin{align}
&\Phi  = \frac{{\left( {1 + k_2^2} \right)\exp \left[ {i{k_1}\sqrt {{{(x - a)}^2} + {y^2} + {z^2}} } \right] - \left( {1 + k_1^2} \right)\exp \left[ {i{k_2}\sqrt {{{(x - a)}^2} + {y^2} + {z^2}} } \right]}}{{2\alpha \sqrt {{{(x - a)}^2} + {y^2} + {z^2}} }}\\
&+\frac{{\left( {1 + k_2^2} \right)\exp \left[ {i{k_1}\sqrt {{{(x + a)}^2} + {y^2} + {z^2}} } \right] - \left( {1 + k_1^2} \right)\exp \left[ {i{k_2}\sqrt {{{(x + a)}^2} + {y^2} + {z^2}} } \right]}}{{2\alpha \sqrt {{{(x + a)}^2} + {y^2} + {z^2}} }},\\
&\Psi = \frac{{\left( {1 + k_1^2} \right)\exp \left[ {i{k_2}\sqrt {{{(x - a)}^2} + {y^2} + {z^2}} } \right] - \left( {1 + k_2^2} \right)\exp \left[ {i{k_1}\sqrt {{{(x - a)}^2} + {y^2} + {z^2}} } \right]}}{{2\alpha \sqrt {{{(x - a)}^2} + {y^2} + {z^2}} }}\\
&+\frac{{\left( {1 + k_1^2} \right)\exp \left[ {i{k_2}\sqrt {{{(x + a)}^2} + {y^2} + {z^2}} } \right] - \left( {1 + k_2^2} \right)\exp \left[ {i{k_1}\sqrt {{{(x + a)}^2} + {y^2} + {z^2}} } \right]}}{{2\alpha \sqrt {{{(x + a)}^2} + {y^2} + {z^2}} }},
\end{align}
\end{subequations}
in which we have taken $Q=1$ and $\Phi_0=\Psi_0=1$ and $\Phi'_0=\Psi'_0=0$, for simplicity. Also, $a$ is the pole spacing. The solution (\ref{db}) represents double quantum interference effect for wavefunction (electron number density) as well as for electrostatic potential field. Using (\ref{mps}) and the standard proceedure it can be shown \cite{akbint} that the total probability current is $J_t=J+J_d$
\begin{subequations}\label{js}
\begin{align}
&J = \frac{i}{2}\left( {{\Psi}\nabla \Psi ^*  - \Psi ^* \nabla {\Psi}} \right),\\
&J_d = - \frac{i}{2}\left( {{\Phi}\nabla \Phi ^*  - \Phi ^* \nabla {\Phi}} \right),
\end{align}
\end{subequations}
where $J$ is the conventional probability current (density) governing all charged and uncharged particles and the new term $J_d$ only rules the charged species. 

\begin{figure}[ptb]\label{Figure3}
\includegraphics[scale=0.75]{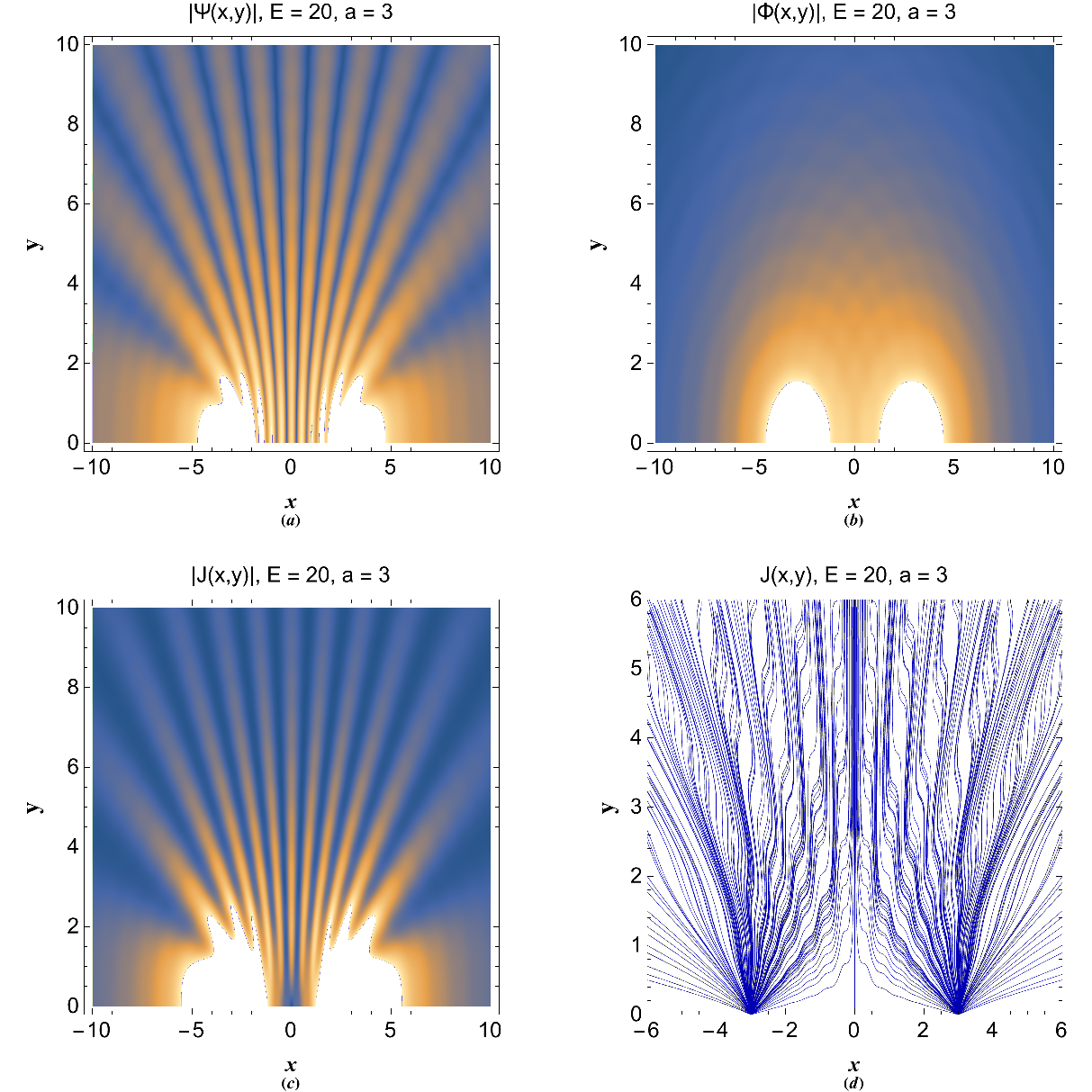}\caption{(a) The dipole interference pattern for quasiparticle wavefunction at orbital $E=20$ with dipole spacing $a=3$. (b) The dipole interference pattern for quasiparticle potential energy disptribution at orbital $E=20$ with dipole spacing $a=3$. (c) The dipole interference probability density for quasiparticle wavefunction at orbital $E=20$ with dipole spacing $a=3$. (d) The deterministic microscopic interference trajectories for quasiparticle wavefunction at orbital $E=20$ with dipole spacing $a=3$.}
\end{figure}

Figure 3 shows the quantum dipole interference which is quite analogous to the well known double-slit phenomenon. We have used $z=0$ in order to study the interference in $x$-$y$ plane. The absolute value of wavefunction is shown at Fig. 3(a) for $E=20$ and $a=3$. The interference pattern shows fringes due to constructive and destructive quantum interferences. In Fig. 3(b) we show the potential field interference for the same parameters. However, while the weak interference pattern is detectable, interference fringes are missing. Figure 3(c) depicts the normal probability current indicating similar pattern as in Fig. 3(a). This figure shows the quantum particle flow intensity in the fluid representation. Figure 3(d) shows the streamplot of the probability current showing the similar pattern as in 3(a) and 3(c). It shows the virtual paths occurring in quantum flow due to dipole interference. This feature is the quite analogous as the pilot-wave theory interpretation of quantum interference effect \cite{de1,de2}.

\begin{figure}[ptb]\label{Figure4}
\includegraphics[scale=0.75]{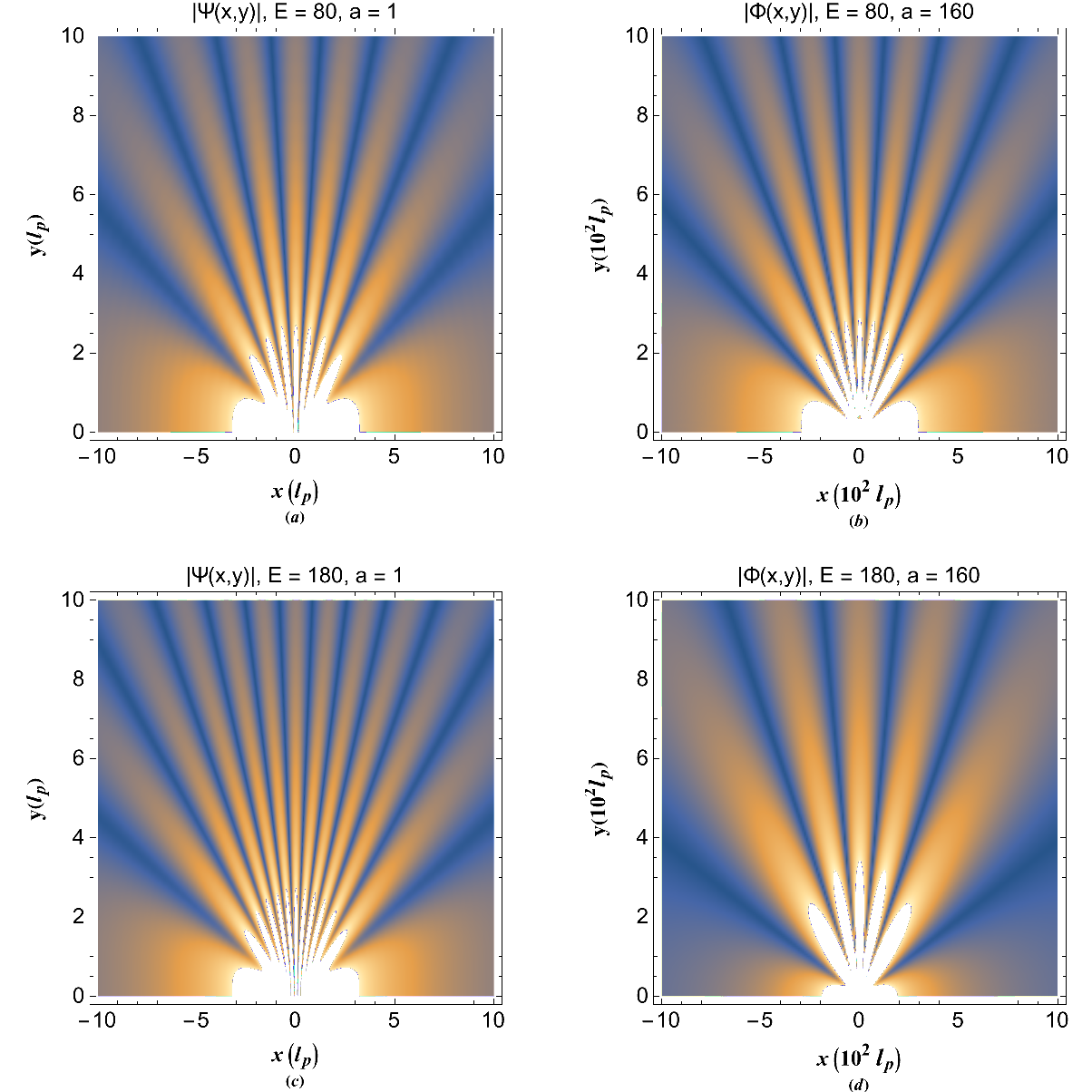}\caption{(a) The dipole interference pattern for quasiparticle wavefunction at orbital $E=80$ with dipole spacing $a=1$ (microscopic level). (b) The dipole interference pattern for quasiparticle potential energy disptribution at orbital $E=80$ with dipole spacing $a=160$ (macroscopic level). (c) The dipole interference pattern for quasiparticle wavefunction at orbital $E=180$ with dipole spacing $a=1$ (microscopic level). (d) The dipole interference pattern for quasiparticle potential energy disptribution at orbital $E=100$ with dipole spacing $a=160$ (macroscopic level).}
\end{figure}

Figure 4 shows the interference patterns for higher quasiparticle energies. Figure 4(a) shows the interference pattern for $|\Psi(x,y)|$ for $E=80$ and $a=1$. Figure 4(b) reveals that fringes appear for $|\Phi(x,y)|$ at this orbital and larger scale ($a=160$). This is expected due to larger de Broglie's wavelength for collective branch compared to that of single-electron one. In Figs. 4(c) and 4(d) we have further increased the quasiparticle energy to see the interference pattern shift. It is remarked that for $E=180$ the spacing between fringes decrease/increase for wavefunction/potential pattern. This feature is also obvious from the energy dispersion curve in Fig. 1(d). It is revealed that, while with increase of quasiparticle orbital energy the wavefunction interference goes microscopic the potential field interference leads to macroscopic effects. 

\begin{figure}[ptb]\label{Figure5}
\includegraphics[scale=0.7]{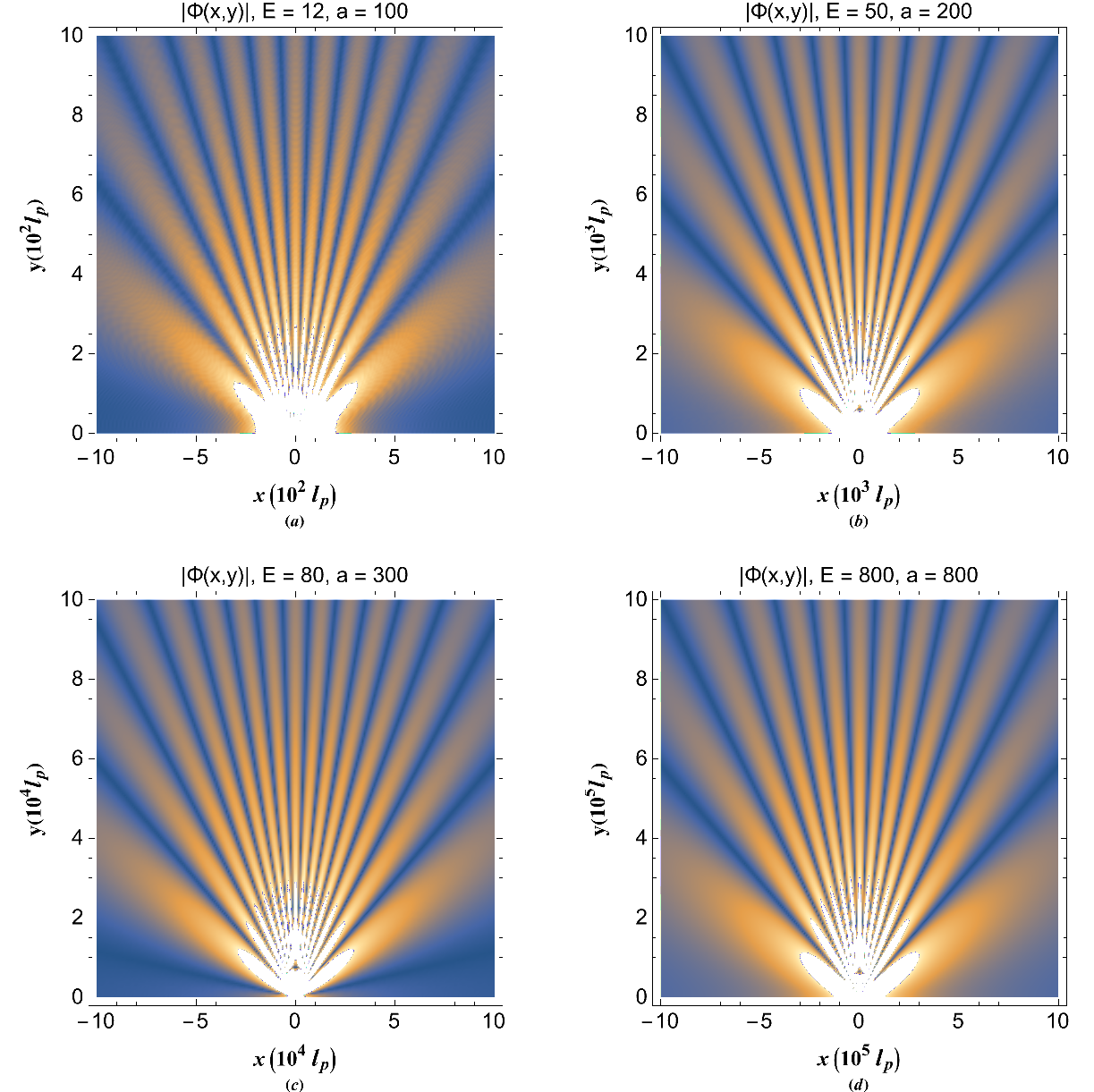}\caption{(a) The dipole interference pattern for quasiparticle potential energy disptribution at orbital $E=12$ with dipole spacing $a=100$ (macroscopic level). (b) The dipole interference pattern for quasiparticle potential energy disptribution at orbital $E=50$ with dipole spacing $a=200$ (macroscopic level). (c) The dipole interference pattern for quasiparticle potential energy disptribution at orbital $E=80$ with dipole spacing $a=300$ (macroscopic level). (d) The dipole interference pattern for quasiparticle potential energy disptribution at orbital $E=800$ with dipole spacing $a=800$ (macroscopic level).}
\end{figure}

The later prediction is confirmed in Fig. 5. In this figure we have consecutively increased the energy and dipole spacings from 5(a) to 5(d) to reach a macroscopic level. It is evident that by proportional increase of these parameters one can arbitrarily increase the macroscopic behavior of the quantum system. This is obviously due to coupling of the macroscopic electrostatic (or other) field to the microscopic quantum effects. As depicted in Fig. 1(a) the characteristic plasmon length varies widely over the electron number density changes. For a typical value of $l_p=10$nm ($n_0\simeq 10^{17}$cm$^{-3}$) the interference lengthscale of Fig. 5(d) falls into the millimeter range. 

\begin{figure}[ptb]\label{Figure6}
\includegraphics[scale=0.7]{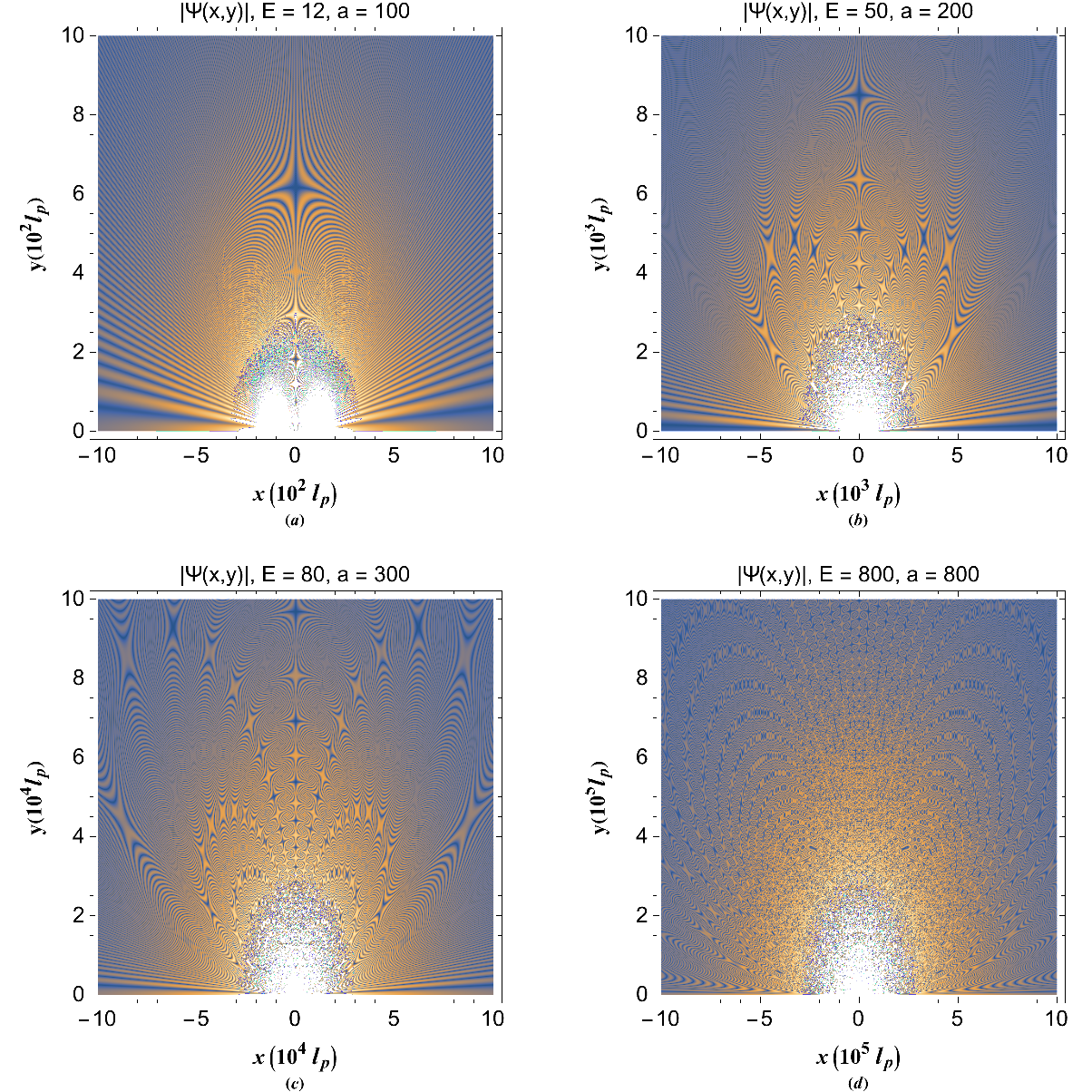}\caption{(a) The dipole interference pattern for quasiparticle wavefunction at orbital $E=12$ with dipole spacing $a=100$ (macroscopic level). (b) The dipole interference pattern for quasiparticle wavefunction at orbital $E=50$ with dipole spacing $a=200$ (macroscopic level). (c) The dipole interference pattern for quasiparticle wavefunction at orbital $E=80$ with dipole spacing $a=300$ (macroscopic level). (d) The dipole interference pattern for quasiparticle wavefunction at orbital $E=800$ with dipole spacing $a=800$ (macroscopic level).}
\end{figure}

In figure 6 we have shown the effects used in Fig. 5 in the case of wavefunction interference. Figure 6(a) reveals an interesting complex wavefunction quantum interference pattern at macroscopic level. Increase of the interference scale parameter in Figs. 6(b) and 6(d) is shown to lead to emergence of fractal-like fine structure in the interference pattern until in Fig. 6(d) the interference pattern fades away at macroscopic level. Therefore, while the potential interference becomes clearer at macroscopic level the wavefunction interference clears away. The later effect is due to the complementarity-like relation $\lambda_1\lambda_2=1$ between the microscopic and macroscopic de Broglie's wavelengths.   

\begin{figure}[ptb]\label{Figure7}
\includegraphics[scale=0.7]{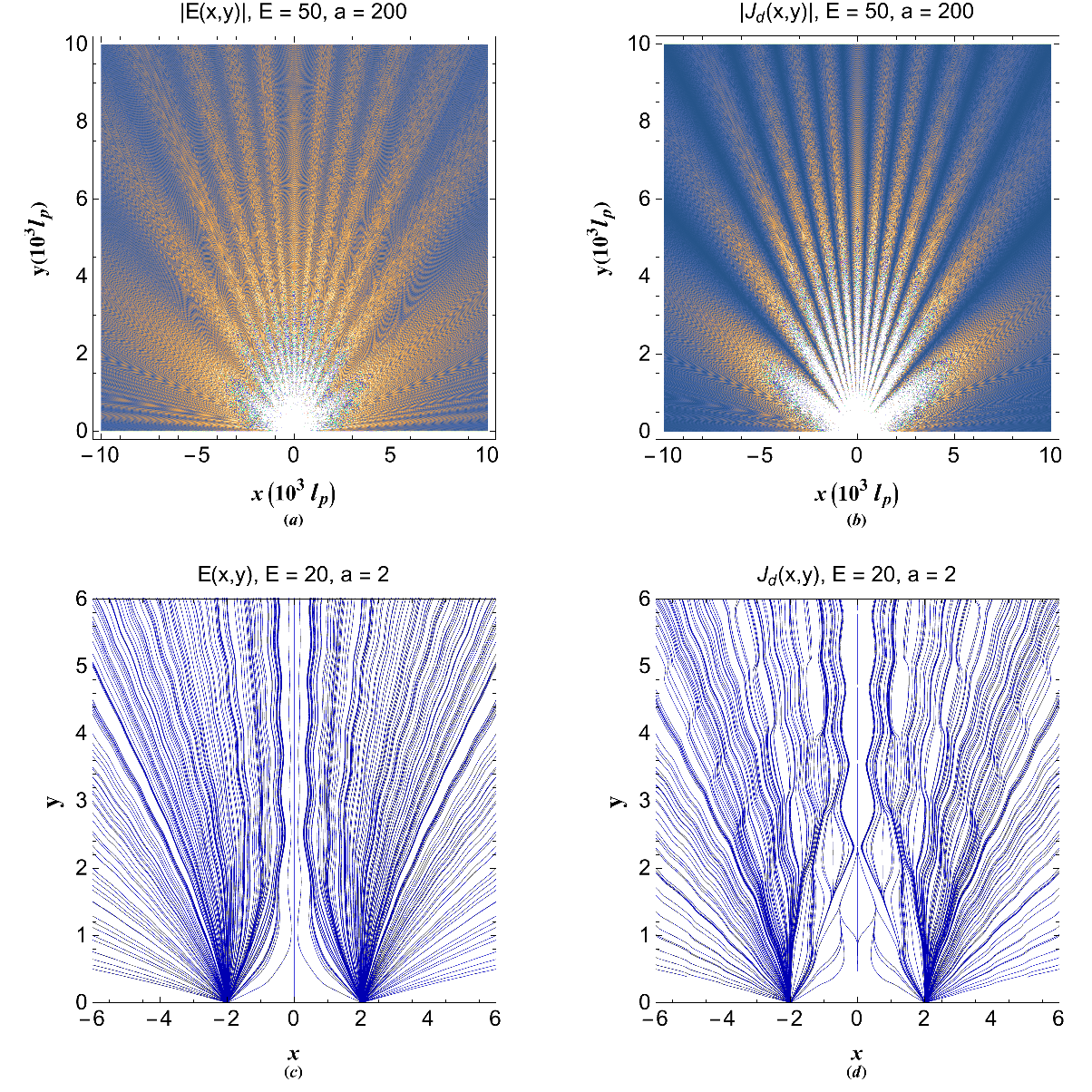}\caption{(a) The dipole interference pattern for electric field distribution at orbital $E=50$ with dipole spacing $a=200$ (macroscopic level). (b) The dipole interference pattern for charge probability density at orbital $E=50$ with dipole spacing $a=200$ (macroscopic level). (c) The streamplot of electric field curves at orbital $E=20$ and $a=2$ (microscopic level). (d) The streamplot of charge current trajectories at orbital $E=20$ and $a=2$ (microscopic level).}
\end{figure}

Figure 7(a) depicts the absolute value of the electric field showing similar interference pattern as in Fig. 5(b). However, there are fine structure modulated over the interference pattern due to the local quantum electron density variations. The presence of such microscopic density pattern makes harder to detect the macroscopic quantum effects. Figure 7(b) shows the probability current part due to the charge transfer as defined in (\ref{js}) for similar interference parameters as in Fig. 7(a). The pattern is quite similar to the electric field interference pattern and predicts a macroscopic effec However, the modulated quantum electron density pattern somehow vagues the vivid interference pattern present in microscopic level. Figures 7(c) and 7(d) show streamplots of electric field and charge probability density revealing the similar structures present in Fig. 3(d) at the microscopic level. 

\section{Conclusion}

Using the quasiparticle model of excitations in arbitrary degenerate electron gas we have shown that macroscopic quantum effects are observable due to dual-tone nature of collective excitations. The generalized energy dispersion of quasiparticle model revealed that many body quantum systems act like a dual arm device acting simultaneously at the microscopic and macroscopic level due to the coupling of quantum field particle interactions to the local particle number density. Current analysis may also have fundamental implications for macroscopic level particle quantum entanglement. This model may be further generalized to include the effects such as electromagnetic interactions and electron spin effects.  

\section{Data Availability}

The data that support the findings of this study are available from the corresponding author upon reasonable request.

\section{Conflict of Interest}

The authors have no conflicts to disclose.

\end{document}